\documentclass[9pt,twocolumn,twoside]{opticajnl}
\journal{opticajournal} 

\setboolean{shortarticle}{true}

\newcommand{\be}{\begin{equation}}
\newcommand{\ee}{\end{equation}}

\title{Guiding light through quasi-TE modes embedded in the radiation continuum}

\author[1,2]{Ji{\v r}{\' i} Petr{\' a}{\v c}ek}
\author[3,*]{Vladim{\' i}r Kuzmiak}
\author[3]{Ji{\v r}{\' i} {\v C}tyrok{\' y}}
\author[4]{Ivan Richter}

\affil[1]{Institute of Physical Engineering, Faculty of Mechanical Engineering, Brno University of Technology, Technick{\'a} 2, 616 69 Brno, Czech Republic}
\affil[2]{Central European Institute of Technology, Brno University of Technology, Purky{\v n}ova, 656/123, 612 00, Brno, Czech Republic}
\affil[3]{Institute of Photonics and Electronics, Academy of Sciences of the Czech Republic, v.v.i., Chabersk{\'a} 57, 182 51 Praha 8, Czech Republic}
\affil[4]{Czech Technical University in Prague, Faculty of Nuclear Sciences and Physical Engineering, Department of Physical Electronics, B{\v r}ehov{\' a} 7, 115 19 Prague 1, Czech Republic}

\affil[*]{Corresponding author: kuzmiak@ufe.cz}

\begin{abstract}
We introduce a new type of a bound state in the continuum (BIC) which appears in the photonic structure consisting of two coupled waveguides where one of them supports a discrete eigenmode spectrum embedded in the continuum of the other one. A BIC appears when the coupling is suppressed by suitable tuning of structural parameters.
In contrast to the previously described configurations, our scheme facilitates genuine guiding of quasi-TE modes in the core with the lower refractive index.
\end{abstract}

\setboolean{displaycopyright}{false}

\begin{document}

\maketitle

Bound states in the continuum (BICs), proposed by von Neumann and Wigner \cite{neumann} in the frame of quantum physics, represent a counterintuitive concept of the confined states with isolated eigenvalues embedded in the continuum of free states. As photonic structures often offer possibilities to tailor the material and structural properties not available in quantum mechanical systems, photonic BICs have become subject of an intensive investigation recently, see, e.g., \cite{Hsu2016,kild}. The most studies on photonic BICs have dealt with periodic structures, such as gratings \cite{Marinica2008}, photonic crystals slabs \cite{Hsu2013}, metasurfaces \cite{Koshelev2018,Koshelev2019}, and waveguide arrays
\cite{plotnik,PK2022b}. Observation of the BICs is fundamentally related to sharp Fano features
in the scattering spectra \cite{Marinica2008,Hsu2013,Koshelev2018,PK2022a}. On the other hand, BICs can also be found in optical waveguides \cite{Zou2015,Nguyen2019}. Such propagating BICs are guided modes with suppressed coupling to the continuum of radiation modes which they are embedded into. Typically, they are formed from quasi-TM leaky modes in waveguides exhibiting the effect of lateral leakage \cite{Oliner1981} when the latter is suppressed due to
destructive interference of radiating TE waves \cite{Webster2007,Nguyen2020}.
The propagating BICs may find an important application in development of new architectures for photonic integrated circuits \cite{Zou2015},
such as the etchless lithium niobate platform \cite{Yu2019,Yu2021}.
Interestingly, properties of several other traditional waveguides, which have been known for decades, can be interpreted in terms of BICs \cite{Ctyroky2023}. Nevertheless, the propagating BICs in waveguides have some limitations:
They are localized in (or in the vicinity of) high-refractive-index materials and, apart from certain types of anisotropic planar waveguides, they are observed only as (quasi) TM polarized waves.
As a result, some of their potential benefits can be disputed \cite{Ctyroky2023,Ctyroky2022}.

In this Letter, we propose a new mechanism of the formation of propagating BICs in systems consisting of
two 
coupled photonic waveguides, where one of them supports a discrete spectrum of guided modes which is embedded in the continuum associated with the other one.
BIC is formed when the coupling between the waveguides is suppressed due to accidental orthogonality of modes in the two individual waveguides; this can be achieved by suitable parametric tuning of the composed structure.
Unlike the previously described designs, our scheme enables quasi-TE-polarized BICs and the modes can be guided in the waveguide core with the lower refractive index.

\begin{figure}[th]
\centering
\includegraphics[width=8cm]{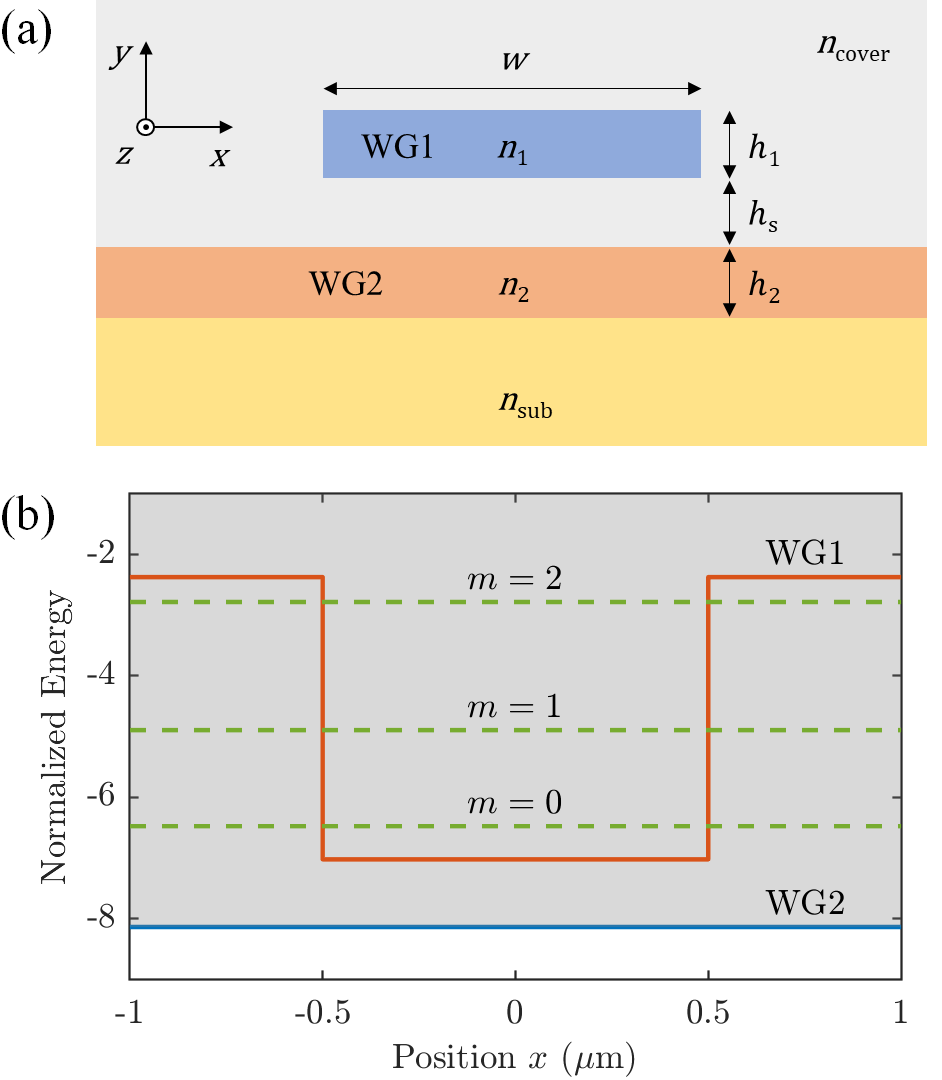}
\caption{(a) The cross-section of the coupled structure with the refractive indices and geometrical parameters indicated.
(b) Potential profiles $V_{1,2}(x)$ corresponding with WG1 and WG2 (solid lines) and energies $E_1$ of bound states (dashed lines) in WG1 labeled with quantum numbers $m$; the continuum of the free states belonging to WG2 is indicated by shaded area.}
\label{fig:Wg_Cross-section}
\end{figure}

Figure~\ref{fig:Wg_Cross-section}(a) shows the proposed structure consisting of a rectangular waveguide WG1 coupled to a planar waveguide WG2. Specifically, we consider a system in SOI platform, where WG2 is made of Si and WG1 is fabricated from InP and is embedded in benzocyclobuten-based (BCB) polymer, which has been proved to be effective in bonding Si wafers with
InP \cite{Maes2012}.
In the calculations we assume the wavelength $\lambda=1.55\,\mu$m and the following structural parameters: $h_1=260\,$nm, $h_2=220\,$nm, $h_{\rm s}=330\,$nm;
$n_1=3.1538$, $n_{\rm cover}=1.54$, $n_2=3.4764$, $n_{\rm sub}=1.44409$.

The standalone WG1 supports quasi-TE$_{m0}$ and quasi-TM$_{m0}$ bound modes, where $m=0,1,2,3\dots$ is the mode order indicating number of nodes in $x$ direction. Propagation constants $k_{1z}$ of the discrete modes are from the range $k_0n_{\rm cover}<k_{1z}<\beta_1$, where $k_0$ is the wavenumber in vacuum and
$\beta_1$ is the propagation constant of the TE$_0$ or TM$_0$ mode propagating along $z$-direction in the {\em planar} waveguide that corresponds to WG1 in the limit of an infinite width, $w\rightarrow \infty$, and which WG1 with finite $w$ is cut from.  
The planar waveguide has the refractive indices $n_1$ (core) and $n_{\rm cover}$ (surrounding) and, for the given core thickness $h_1$, is single mode. The standalone WG2, i.e., the planar waveguide with the refractive 
indices $n_2$ (core), 
$n_{\rm sub}$ (substrate) and $n_{\rm cover}$ (cover), supports the continuum of the radiation modes propagating in 
$xz$ plane with the propagation constants
$k_{2z}\le \beta_2$, where $\beta_2$ is the propagation constant of
the relevant mode (TE$_0$ or TM$_0$) propagating along $z$-direction in the waveguide. 
Hereafter, we use the subscripts 1 and 2 to denote quantities belonging to WG1 and WG2, respectively.

By using the isomorphism of the eigenvalue equation for $N_{{\rm eff},j}$, where $N_{{\rm eff},j} \equiv k_{jz}/k_0$ is the effective index of modes supported by WG$j$, $j=1,2$, and the steady-state Schr{\"o}dinger equation, 
we can construct the potential profiles,
see Fig.~\ref{fig:Wg_Cross-section}(b),
describing the quantum mechanical analogue of the proposed photonic structure.
Here WG1 corresponds to a quantum well and WG2 is described by a constant potential.
We use normalized quantities, therefore the potential $V_1(x)$ in WG1 corresponds with the effective refractive index profile
$n_{\rm eff}(x)$ obtained from the effective index method \cite{Kogelnik1990} as $V_1(x)=-n_{\rm eff}^2(x)$ and the energy eigenvalues
are given as $E_1=-N_{{\rm eff},1}^2$. Likewise, the potential $V_2(x)$ in WG2 is calculated as $V_2(x)=-(\beta_2/k_0)^2$.
Explicitly, we considered only quasi-TE modes and assumed WG1 with width $w=1.0\,\mu$m.
Fig.~\ref{fig:Wg_Cross-section}(b) demonstrates that the bound states associated with WG1 are embedded in the continuum of the free states belonging to WG2. Therefore, in the composed structure, the bound modes couple to the continuum and become leaky. However,
by suitable tuning the width $w$, the coupling can be suppressed and the leaky modes turn into BICs.

Qualitatively, we can assume in either waveguide that modal fields oscillate in $x$
with spatial frequencies $k_{jx}$, $j=1,2$, that satisfy the following relations
\be \label{eq_dispersion}
k_{jx}^2+k_{jz}^2=\beta^2_{j}.
\ee
In WG1 the oscillations occur only in the range $-w/2<x<w/2$ that defines the horizontal extent of the InP core.
In the composed structure, the modes of WG1 and WG2 couple to each other, however, their interaction is efficient only  when they are phase matched, i.e., $k_{2z}=k_{1z}$, which yields
\be \label{eq_kx_exact}
k_{2x} = \sqrt {\beta_2^2 - \beta_1^2 + k_{1x}^2 }.
\ee
On the other hand, in WG1, for an arbitrary mode far from its cutoff, we can estimate $k_{1x}w\approx (m+1)\pi$.
Furthermore, the coupling, which occurs in the range $-w/2<x<w/2$, is suppressed
when  $k_{2x}w -k_{1x}w =2p\pi$, $p=1,2,3\dots$, i.e., when the oscillating profiles are orthogonal.
By using these conditions in Eq.~(\ref{eq_kx_exact}) one obtains the expression
\be \label{eq_w_BIC}
w_{\rm BIC} \approx \frac{2\pi}{\sqrt{\beta_2^2-\beta_1^2}}\sqrt{p^2+p(m+1)},
\ee
which predicts the critical widths of WG1, $w\equiv w_{\rm BIC}$, for which the composed structure supports BICs.

\begin{figure}[!htb]
\centering
\includegraphics[width=8cm]{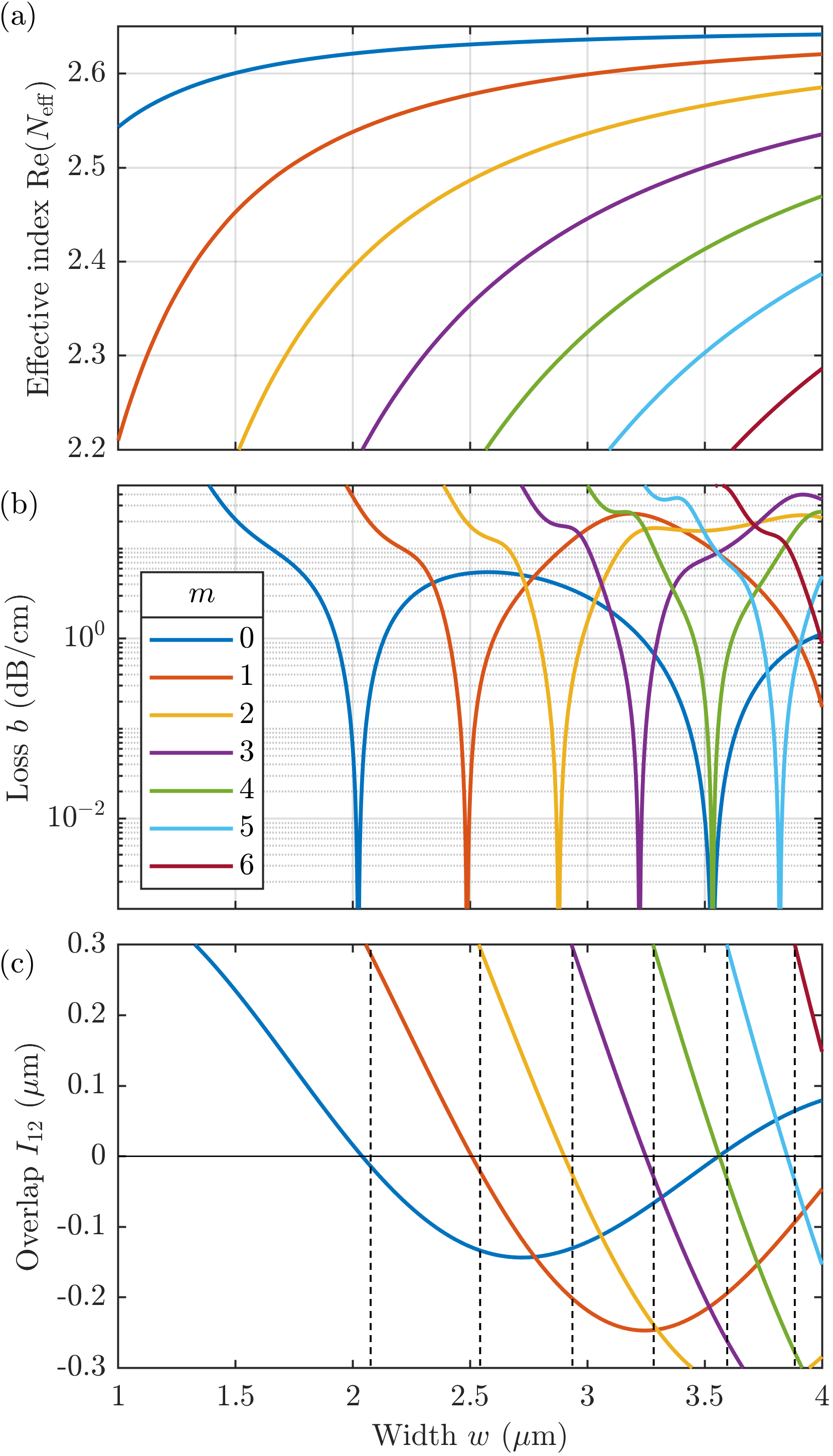}
\caption{(a), (b) Leaky modes of the composed structure: (a) Real part of the mode effective index $N_{\rm eff}$ and (b) loss $b$ vs width
$w$ of InP waveguide; only the modes corresponding with quasi-TE$_{m0}$ modes of WG1 are shown, the mode order $m$ is specified in (b).	
(c) Corresponding dependencies of $I_{12}(w)$, vertical dashed lines indicate positions of BICs predicted through Eq.~(\ref{eq_w_BIC}).
}
\label{fig:Neff}
\end{figure}

To verify the predictions of our simplified model and to prove the existence of BICs,
we rigorously calculated supermodes of composed structures with various widths $w$ by using COMSOL.
While the structure exhibits a rich spectral behavior, in presenting the results in Fig.~\ref{fig:Neff}(a) and (b)
we restrict ourselves to leaky modes that correspond to quasi-TE$_{m0}$ modes of WG1.
The real components of their effective indices behave as expected for the modes of WG1, however the presence of WG2 induces a loss, which strongly depends on the width $w$.
One can see that at certain ``magic'' widths, the leakage vanishes and a BIC is formed.
The positions of the magic widths observed in Fig.~\ref{fig:Neff}(b) are in accord
with values predicted by Eq.~(\ref{eq_w_BIC}), which are indicated by vertical dashed lines in Fig.~\ref{fig:Neff}(c).
We note that Eq.~(\ref{eq_w_BIC}) predicts existence of two BICs for the same structure, e.g., BICs with $m=0,\,p=2$ and $m=4,\,p=1$ are found simultaneously
for $w_{\rm BIC} \approx 3.6\,\mu$m; as it is shown in Fig.~\ref{fig:Neff}(b).

\begin{figure}[!htb]
\centering
\includegraphics[width=8.7cm]{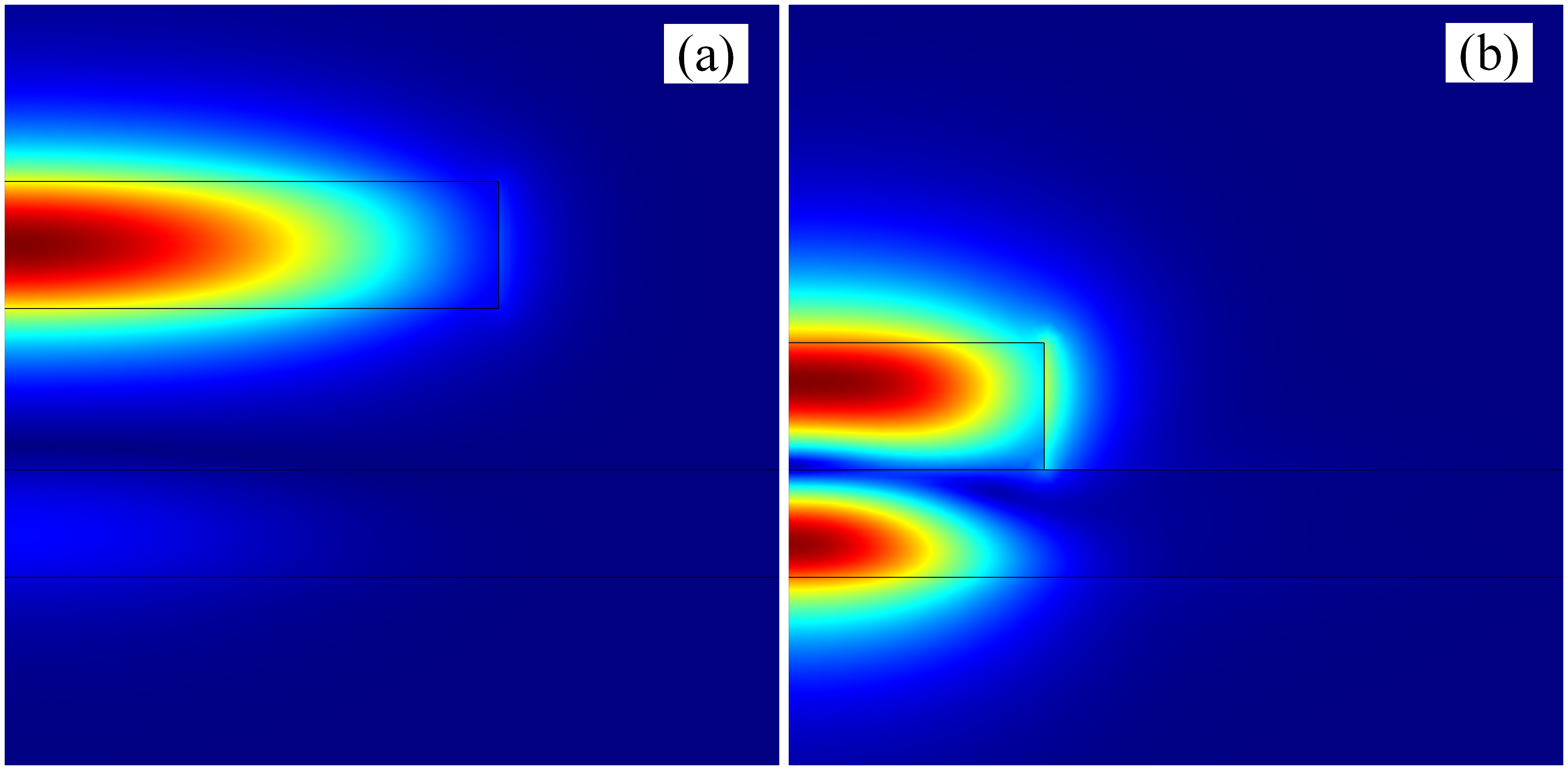}
\caption{Electric field profile $|\vec E|$ for the quasi-TE BIC defined with $m=0$, $p=1$ for structures with different separation distances $h_s$.
(a) $h_s=330\,$nm, the BIC corresponds to minima at $w=2.023\,\mu$m in Fig.~\ref{fig:Neff}(b).
(b) $h_s=0$, the BIC occurs at $w=1.047\,\mu$m, see Fig.~\ref{fig:effectofhs}.
Only symmetric half of the profile is shown.}
\label{fig:ModalField}
\end{figure}

The localized pattern of the field associated with BICs confined in WG1, i.e., in the core with the lower refractive index, is illustrated in Fig.~\ref{fig:ModalField}(a).

The simplified model leading to Eq.~(\ref{eq_w_BIC}) can be cast in a more rigorous form using the coupled mode theory (CMT).
We start with modal fields $\vec E_j(x,y,z,t) = \vec {\cal E}_j(x,y) \exp(i\omega t -i k_{jz}z)$ in the individual waveguides. Their transversal profiles that describe quasi-TE modes have the form
\begin{align}
{\cal E}_{jx}(x,y)&=C_j\varphi_j(y)\psi_j(x)  												\label{eq_mode_Ex} \\
{\cal E}_{jy}(x,y)&=0  																								\label{eq_mode_Ey} \\
{\cal E}_{jz}(x,y)&=-i\frac{C_j}{k_{jz}}\varphi_j(y)\psi_j^\prime(x) 	\label{eq_mode_Ez} \\
 \psi_j(x)&\equiv \cos(k_{jx}x-\alpha_j)  														\label{eq_def_psi}
\end{align}
where $C_j$ is a normalization constant and $\varphi_j(y)$ describes the dependence on $y$. Eqs.~(\ref{eq_mode_Ex})-(\ref{eq_def_psi}) supplied with Eq.~(\ref{eq_dispersion})
are exact for modes in WG2 ($j=2$) while only approximate for WG1 ($j=1$). The approximation is consistent with the Marcatili’s technique \cite{Marcatili1969} in the formulation that assures $E_y=0$
\cite{Westerveld2012}, which is used here for determination of $k_{1x}$ and $k_{1z}$ in the subsequent CMT calculation. Indeed, $k_{2x}$ follows from Eq.~(\ref{eq_kx_exact}) and $k_{2z}=k_{1z}$.
The parameter $\alpha_j$ in Eq.~(\ref{eq_def_psi}) describes symmetry of a mode with respect to $yz$ plane. For WG1, $\alpha_1$ equals either $0$ or $\pi$ when $m$ is even or odd, respectively.
Radiation modes of WG2 are doubly-degenerate with $\alpha_2=0$ and $\alpha_2=\pi$ for a given propagation constant $k_{2z}$; however, only one mode from the degenerate mode pair has a suitable symmetry enabling interaction with a given mode of WG1, thus we can set $\alpha_2=\alpha_1$.

Upon suitable normalization of the modal fields, the coefficients describing the coupling read according to \cite{Marcuse1972}
\be \label{eq_K_jk}
K_{jk}=k_0\iint \left(\varepsilon - \varepsilon_k\right)
\vec{\mathcal{E}}_{j}^- \cdot \vec{\mathcal{E}}_{k} \, dxdy,
\ee
where $j \neq k$ ($j,k=1,2$), $\varepsilon$ is the relative permittivity profile describing the whole structure, $\varepsilon_k$ is the profile used
for calculation of the mode labeled with $k$ and $\vec{\mathcal{E}}_{j}^-$ results from $\vec{\mathcal{E}}_{j}$ by changing the sign of $k_{jz}$.
One can see from Eqs.~(\ref{eq_mode_Ex})-(\ref{eq_def_psi}) that the integration in $x$ and $y$ in Eq.~(\ref{eq_K_jk}) can be separated and as a result $K_{jk}$ is directly proportional to the overlap
\be \label{eq_I_jk}
I_{jk} = \int \limits_{-w/2}^{w/2} \left[\psi_j(x) \psi_k(x)  + \frac{1}{k_{jz}k_{kz}}\psi_j^\prime(x) \psi_k^\prime(x) \right]dx.
\ee
Analytical calculation of the integral in Eq.~(\ref{eq_I_jk}) is straightforward. As $I_{12}=I_{21}$, roots of the dependence $I_{12}(w)$ provide magic widths $w_{\rm BIC}$ leading to the inhibited coupling,
$K_{12}=0$, $K_{21}=0$, and, consequently the existence of BIC.

Figure~\ref{fig:Neff}(c) displays dependencies $I_{12}(w)$ for modes of WG1 with various orders $m$. The positions of the roots agree very well with locations of BICs in
Fig.~\ref{fig:Neff}(b).

So far, in our models 
we neglected the effect of separation distance $h_{\rm s}$. This is justified for the geometry studied in
Fig.~\ref{fig:Neff}. However, when $h_{\rm s}$ decreases, the presence of one waveguide can affect modes of the other one and significantly alter magic widths.
To take this effect into account we calculate the two supermodes of the {\em planar} structure corresponding to the composed structure in Fig.~\ref{fig:Wg_Cross-section} in the limit $w\rightarrow\infty$ and identify their propagation constants with $\beta_1$ and $\beta_2$ in Eqs.~(\ref{eq_dispersion}) and (\ref{eq_kx_exact}) to obtain the parameters needed
for evaluation of Eq.~(\ref{eq_I_jk}). Throughout the calculation $k_{1x}$ is considered to be independent
of $h_{\rm s}$, this is reasonable because $k_{1x}$ depends mainly on $w$.

\begin{figure}[tbp]
\centering
\includegraphics[width=8cm]{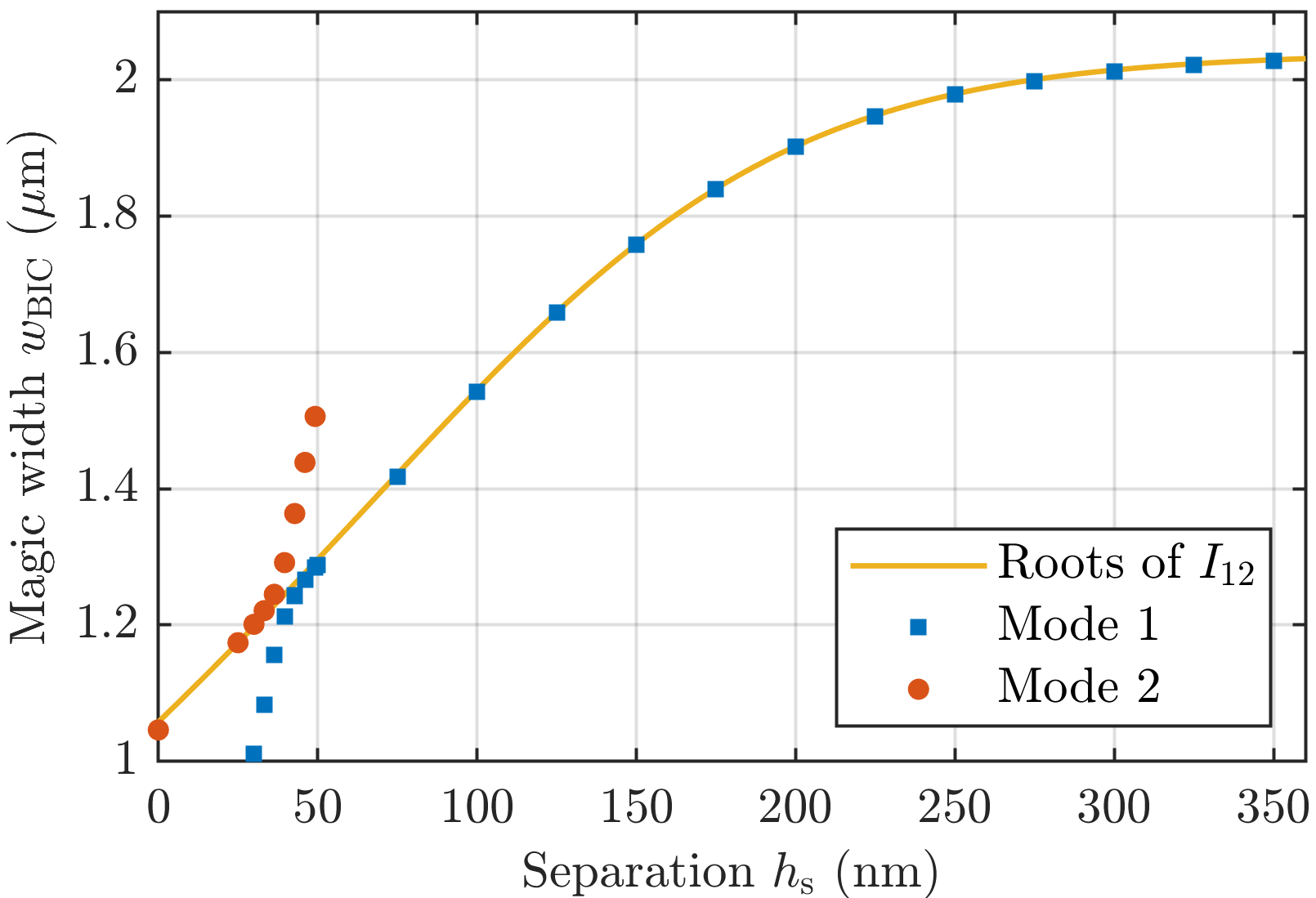}
\caption{Effect of the separation distance $h_{\rm s}$ on the magic width $w_{\rm BIC}$ for the BIC corresponding to minima at $w=2.023\,\mu$m (i.e., $m=0$, $p=1$) in Fig.~\ref{fig:Neff}(b).}
\label{fig:effectofhs}
\end{figure}

The effect of the separation distance $h_{\rm s}$ on the magic width $w_{\rm BIC}$ is demonstrated in Fig.~\ref{fig:effectofhs}. For the sake of conciseness, we considered only the BIC with $m=0$ and $p=1$. The CMT model (the solid line) shows that decreasing $h_{\rm s}$ leads to narrowing
$w_{\rm BIC}$. This agrees with the prediction of Eq.~(\ref{eq_w_BIC}): as the difference $\beta_2-\beta_1$ increases with decreasing $h_{\rm s}$, so does the denominator on the r.h.s. of Eq.~(\ref{eq_w_BIC}). Fig.~\ref{fig:effectofhs} also presents results of rigorous COMSOL calculations, which reveal two modes (Mode 1 and Mode 2), which exhibit the polarization conversion.
In the limit of large $h_{\rm s}$, Mode 1 coincides with the investigated (quasi-TE polarized) BIC. The behavior of Mode 2 in this limit is more complicated; however, for $h_{\rm s}\approx 50\,$nm, Mode 2 describes a certain quasi-TM polarized BIC.
In the anticrossing region, near $h_{\rm s}\approx 40\,$nm, both the modes become strongly hybrid. With further decreasing of
$h_{\rm s}$ the polarization of the modes is reversed and the studied quasi-TE BIC coincides with Mode 2.
The CMT model cannot describe the avoided crossing, as we did not assumed the interaction (the study of the quasi-TM BICs and their interactions with quasi-TE modes is out of scope of this Letter).
However, apart from this narrow region, the CMT model is in perfect agreement with the rigorous calculations.
Simulations also reveal that with decreasing $h_{\rm s}$ the BIC ceases to be confined in WG1, as illustrated with the field distribution for
$h_{\rm s}=0$ in Fig.~\ref{fig:ModalField}(b).

In summary, we theoretically proposed a new principle which introduces an additional degree of freedom to existing approaches of formation of the BICs in the photonic structures. The concept was
demonstrated in a coupled waveguide structure consisting of a rectangular and planar waveguide where bound modes of the one waveguide are embedded in the continuum of the other one.
We developed a simple qualitative model that captures the underlying physics: the coupling between the bound and radiation modes can be suppressed by parametric tuning resulting in BIC formation.
Combining these ideas with the coupled mode theory provides deeper insight and a general framework for arbitrary
parametric study. Here we demonstrated the effect of width of the rectangular waveguide and influence of the separation distance. These predictions were confirmed numerically by rigorous numerical calculations by using COMSOL. We believe that this approach expands a rich variety of applications associated with BICs due to its simplicity and may offer new possibilities of creation of the propagating BICs.

\begin{backmatter}

\bmsection{Disclosures}
The authors declare no conflicts of interest.

\end{backmatter}




\begin{thebibliography}{99}
\newcommand{\enquote}[1]{``#1''}

\bibitem{neumann} J. von Neuman and E. Wigner,
\enquote{\"{U}ber merkw\"{u}rdige dikrete Eigenwerte},
Phys. Z. {\bf 30}, 465-467 (1929).

\bibitem{Hsu2016}
C. W Hsu, B. Zhen, A. D. Stone, J. D. Joannopoulos, and M. Solja{\v c}i\'c,
 \enquote{Bound states in the continuum},
Nat. Rev. Mater. {\bf 1}, 16048 (2016).

\bibitem{kild} S. I. Azzam and A. V. Kildishev,
 \enquote{Photonic Bound States in the Continuum: From Basics to Applications},
Adv. Opt. Mater. {\bf 9}, 2001469 (2021).

\bibitem{Marinica2008}
D. C. Marinica, A. G. Borisov, and S. V. Shabanov,
 “Bound states in the continuum in photonics”,
Phys. Rev. Lett. {\bf 100}, 183902 (2008).

\bibitem{Hsu2013}
C. W. Hsu, B. Zhen, J. Lee, S.-L. Chua, S. G. Johnson, J. D. Joannopoulos, and M. Solja{\v c}i\'c,
“Observation of trapped light within the radiation continuum”,
Nature {\bf 499}, 188–191 (2013).

\bibitem{Koshelev2018}
K. Koshelev, S. Lepeshov, M. Liu, A. Bogdanov, and Y. Kivshar,
 “Asymmetric Metasurfaces with High- Q Resonances Governed by Bound States in the Continuum”,
Phys. Rev. Lett. {\bf 121}, 193903 (2018).

\bibitem{Koshelev2019}
K. Koshelev, Y. Tang, K. Li, D.-Y. Choi, G. Li, and Y. Kivshar,
 “Nonlinear Metasurfaces Governed by Bound States in the Continuum”,
ACS Photonics {\bf 6}, 1639–1644 (2019).

\bibitem{plotnik}
Y. Plotnik, O. Peleg, F. Dreisow, M. Heinrich, S. Nolte, A. Szameit, and M. Segev,
\enquote{Experimental Observation of Opical Bound States in the Continuum},
Phys. Rev. Lett. {\bf 107}, 183901 (2011).

\bibitem{PK2022b} J. Petr{\' a}{\v c}ek and V. Kuzmiak,
"Bound states in the continuum in waveguide arrays within a symmetry classification scheme",
Opt. Express {\bf 30}, 35712-35724 (2022).

\bibitem{PK2022a} J. Petr{\' a}{\v c}ek and V. Kuzmiak,
 \enquote{Effect of symmetry breaking on bound states in the continuum in waveguide arrays},
Phys. Rev. A {\bf 105}, 063505 (2022).

\bibitem{Zou2015} C.-L. Zou, J.-M. Cui, F.-W. Sun, X. Xiong, X.-B. Zou, Z.-F. Han, and G.-C. Guo,
"Guiding light through optical bound states in the continuum for ultrahigh-Q microresonators",
Laser Photonics Rev. {\bf 9}, 114-119 (2015).

\bibitem{Nguyen2019} T. G. Nguyen, G. Ren, S. Schoenhardt, M. Knoerzer, A. Boes, and A. Mitchell,
"Ridge Resonance in Silicon Photonics Harnessing Bound States in the Continuum",
Laser Photonics Rev. {\bf 13}, 1900035 (2019).

\bibitem{Oliner1981} A. A. Oliner, S. T. Peng, T. I. Hsu, and A. Sanchez,
"Guidance and leakage properties of a class of open dielectric wave-guides .2. New physical effects",
IEEE Trans. Microwave Theory Tech. {\bf 29}, 855-869 (1981).

\bibitem{Webster2007} M. A. Webster, R. M. Pafchek, A. Mitchell, and T. L. Koch,
 “Width dependence of inherent TM-mode lateral leakage loss in silicon-on-insulator ridge waveguides”,
IEEE Photonics Technol. Lett. {\bf 19}, 429–431 (2007).

\bibitem{Nguyen2020} T. G. Nguyen, A. Boes, and A. Mitchell,
"Lateral Leakage in Silicon Photonics: Theory, Applications, and Future Directions",
IEEE J. Sel. Top. Quantum Electron. {\bf 26}, 8200313 (2020).

\bibitem{Yu2019}  Z. Yu, X. Xi, J. Ma, H. K. Tsang, C.-L. Zou, and X. Sun,
 "Photonic integrated circuits with bound states in the continuum",
Optica {\bf 6}, 1342-1348 (2019).

\bibitem{Yu2021}
Y. Yu, Z. Yu, L. Wang, and X. Sun,
 “Ultralow-Loss Etchless Lithium Niobate Integrated Photonics at Near-Visible Wavelengths,”
Adv. Opt. Mater. {\bf 9}, 2100060 (2021).

\bibitem{Ctyroky2023}
J. {\v C}tyrok{\' y}, J. Petr{\' a}{\v c}ek, V. Kuzmiak, and I. Richter,
"Bound modes in the continuum in integrated photonic LiNbO3 waveguides: are they always beneficial?",
Opt. Express {\bf 31}, 44-55 (2023).

\bibitem{Ctyroky2022}
J. {\v C}tyrok{\' y} and J. Petr{\' a}{\v c}ek,
“Comment on “Photonic integrated circuits with bound states in the continuum”,
Optica {\bf 9}, 681-682 (2022).

\bibitem{Maes2012}
B. Maes, J. Petr{\' a}{\v c}ek, S. Burger, P. Kwiecien, J. Luksch, and I. Richter,
 "Simulations of high-Q optical nanocavities with a gradual 1D bandgap",
Opt. Express {\bf 21}, 6794-6806 (2013).

\bibitem{Kogelnik1990}
H. Kogelnik,
 Theory of Optical Waveguides,
in {\em Guided-Wave Optoelectronics}, T. Tamir, ed., Vol. 26 of Springer Series in Electronics and Photonics, 2nd ed. (Springer, 1990),
Chap. 2, p. 69.

\bibitem{Marcatili1969}
E. Marcatili,
 "Dielectric rectangular waveguide and directional coupler for integrated optics",
Bell Syst. Tech. J. {\bf  48}, 2071–2121 (1969).

\bibitem{Westerveld2012}
W. J. Westerveld, S. M. Leinders, K. W. A. Van Dongen, H. P. Urbach, and M. Yousefi,
 "Extension of Marcatili’s analytical approach for rectangular silicon optical waveguides",
J. Light. Technol. {\bf 30}, 2388–2401 (2012).

\bibitem{Marcuse1972}
D. Marcuse, {\em Light Transmission Optics} 
(Van Nostrand, 1972).
\end{thebibliography}
\end{document}